\documentclass{PoS}

\def\tr{{\mbox{tr}}}
\def\omegabar{{\overline{\omega}}}
\def\chibar{{\overline{\chi}}}
\def\lambdabar{{\overline{\lambda}}}
\def\psibar{{\overline{\psi}}}

\title{Vacuum alignment and lattice artifacts}

\ShortTitle{Vacuum alignment and lattice artifacts}

\author{\speaker{Maarten Golterman}%
        \null\\
        Department of Physics and Astronomy, San Francisco State University, \\
        San Francisco, CA 94132, USA\\
        E-mail: \email{maarten@sfsu.edu}}

\author{Yigal Shamir\\
        Raymond and Beverly Sackler School of Physics and Astronomy,
Tel Aviv University, Ramat~Aviv,~69978~ISRAEL\\
        E-mail: \email{shamir@post.tau.ac.il}}

\abstract{When a subgroup of the flavor symmetry group of a gauge theory is weakly coupled to additional gauge fields, the vacuum tends to align such that the gauged subgroup is unbroken. At the same time, the lattice discretization typically breaks the flavor symmetry explicitly, and can give rise to new lattice-artifact phases with spontaneously broken symmetries. We discuss the interplay of these two phenomena, using chiral lagrangian techniques. Our  first example is two-flavor Wilson QCD coupled to electromagnetism. We also consider examples of theories with staggered fermions, and demonstrate that recent claims
in the literature based on the use of staggered fermions are incorrect.}

\FullConference{The 32nd International Symposium on Lattice Field Theory\\
                 23-28 June, 2014\\
                 Columbia University New York, NY}

\begin{document}

\section{Overview}
In two recent articles, we considered the question of what happens if, in a strongly-coupled, vector-like lattice
gauge theory with a global flavor symmetry, a subgroup of this flavor symmetry is weakly gauged
\cite{Golterman:2014yha,Golterman:2014lea}.  We consider the flavor gauge coupling to be ``weak'' if it is small at the scale of the strongly-coupled theory.   In the continuum limit of such theories, the
vacuum tends to align such that the weakly gauged subgroup remains unbroken \cite{MP1980}.
However, on the lattice, even without gauging any of the flavor symmetries, it may happen that
flavor symmetries are explicitly and/or spontaneously broken because of lattice artifacts.   It follows that in
lattice gauge theories with weakly gauged flavor symmetries, there may be a competition between
vacuum alignment and lattice-induced symmetry breaking, and it is interesting to consider in more
detail how this competition plays out.

Here we consider this question for two examples, the first being QCD with two flavors of Wilson
fermions, and the second QCD with two flavors of (unrooted) staggered fermions.   For other
examples, we refer to Refs.~\cite{Golterman:2014yha,Golterman:2014lea}.   The upshot is that
indeed lattice artifacts can change the continuum phase diagram in a non-trivial way, as we will
see in the examples below.   We will begin with a quick review of two-flavor QCD coupled to isospin
gauge fields
in the continuum.  We then consider lattice QCD with two flavors of Wilson fermions in Sec.~3, and
with two flavors of staggered fermions in Sec.~4, after which we conclude in the final section.

\section{Continuum case}
The low-energy effective potential for two-flavor QCD with degenerate quark masses is given
by
\begin{eqnarray}
\label{Veff}
V_{\rm eff}&=&-\frac{c_1}{4}\,\tr(\Sigma+\Sigma^\dagger)\ ,\\
\Sigma&=&\sigma+i{\vec\tau}\cdot{\vec\pi}\ ,\qquad\sigma^2+{\vec\pi}^2=1\ ,\nonumber
\end{eqnarray}
in which the constant $c_1$ is proportional to the quark mass, and $\Sigma$ is the usual
$SU(2)$-valued non-linear pion field.   Instead of using the standard exponential form for
$\Sigma$, we employ the parametrization shown in Eq.~(\ref{Veff}), as it is more
convenient in the case of $SU(2)$.
We may gauge isospin (or a subgroup thereof)
by considering the kinetic term of the effective lagrangian, and replacing the derivatives
by covariant derivatives,
\begin{equation}
\label{kin}
\frac{f^2}{8}\,\tr(\partial_\mu\Sigma\partial_\mu\Sigma^\dagger)\rightarrow
\frac{f^2}{8}\,\tr(D_\mu\Sigma(D_\mu\Sigma)^\dagger)\ ,
\end{equation}
with the covariant derivative
\begin{equation}
\label{covD}
D_\mu\Sigma=\partial_\mu\Sigma+ig[V_\mu,\Sigma]\ ,\qquad V_\mu={\vec V}_\mu\cdot{\vec\tau}/2\ ,
\end{equation}
with $V_\mu$ the isospin gauge field.   From Eq.~(\ref{kin}) one then reads off the non-derivative
part
\begin{equation}
\label{nonderiv}
\frac{g^2f^2}{4}\,\tr(V_\mu^2-V_\mu\Sigma V_\mu\Sigma^\dagger)\ ,
\end{equation}
and, integrating over $V_\mu$, this leads to an addition to the effective potential
\begin{equation}
\label{deltaVeff}
\Delta V_{\rm eff}=-\frac{g^2c_3}{8}\,\sum_a\tr(\tau_a\Sigma\tau_a\Sigma^\dagger)=-g^2c_3\sigma\ ,
\end{equation}
to lowest non-trivial order in the weak coupling $g$.   We may also consider the case that we couple
only a $U(1)$ subgroup of isospin $SU(2)$.   Setting $V^1_\mu=V^2_\mu=0$, {\it i.e.}, taking this $U(1)$ in the $\tau_3$ direction, one finds
that
\begin{eqnarray}
\label{VeffEM}
V_{\rm eff}+\Delta V_{\rm eff}&=&-\frac{c_1}{4}\,\tr(\Sigma+\Sigma^\dagger)
-\frac{e^2c_3}{8}\,\tr(\tau_3\Sigma\tau_3\Sigma^\dagger)\\
&=&\mbox{constant}+\frac{1}{2}\,c_1\,{\vec\pi}^2+e^2c_3\,\pi^+\pi^-+\dots\ .\nonumber
\end{eqnarray}
We thus rederive the ancient result \cite{EMpion}
\begin{equation}
\label{pimassdiff}
m^2_{\pi^+}-m^2_{\pi^0}=e^2c_3/f^2\ ,
\end{equation}
in which we have that $c_3>0$ \cite{EW1983}.  We note that in both cases gauging
(a subgroup of) the isospin group
stabilizes the vacuum, because $c_3$ is positive.   This is an example of vacuum
alignment \cite{MP1980}.

\section{Wilson fermions}
In the regime in which $m/\Lambda_{\rm QCD}\sim a^2\Lambda_{\rm QCD}^2$, the
leading-order effective potential for QCD with two flavors of Wilson fermion is \cite{ShSi1998}
\begin{eqnarray}
\label{VeffWilson}
V_{\rm eff}&=&-\frac{c_1}{4}\,\tr(\Sigma+\Sigma^\dagger)+\frac{c_2}{16}\,\left(\tr(\Sigma+\Sigma^\dagger)\right)^2\\
&=&-c_1\sigma+c_2\sigma^2\ ,\nonumber
\end{eqnarray}
where $c_1\propto m$ as before,\footnote{The term of order $a$ in the effective potential can be, and thus has been, absorbed into $m$.  For
details on how this works, see Ref.~\cite{ShSi1998}.} and $c_2\propto a^2$, with $a$ the lattice spacing.

If $c_2<0$, then $\langle\sigma\rangle=\pm 1$, depending on the sign of $c_1$.  There is a first-order
phase transition at $c_1=0$, and isospin is always unbroken.   In this case, the minimal value of the pion mass is proportional to $|c_2|$;
no pion is ever massless as long as the lattice spacing $a>0$.

If $c_2>0$, one finds that the effective potential is minimized at
\begin{equation}
\label{Vmin}
\langle\sigma\rangle=
\left\{\begin{array}{ll}
1\ ,&\ \ c_1\ge 2c_2\ ,\\
\frac{c_1}{2c_2}\ ,&\ \ -2c_2<c_1<2c_2\ ,\\
-1\ ,&\ \ c_1\le -2c_2\ .
\end{array}\right.
\end{equation}
There is a second-order phase transition at $|c_1|=2c_2$, and
for $|c_1|<2c_2$ we find that $|\langle\sigma\rangle|<1$, which implies that $\langle\vec\pi\rangle\ne 0$.   $SU(2)$ isospin is spontaneously broken to a $U(1)$
subgroup, parity is spontanteously broken, and two of the three pions become massless as the Nambu--Goldstone bosons associated
with this symmetry breaking.  This region in the phase diagram is the
Aoki phase \cite{Aoki1983}.   We note that the effect on pion masses is opposite to the effect of
coupling QCD to QED, with the neutral pion being heavier than the charged pion inside the Aoki phase!%
\footnote{Here the neutral pion is by definition the excitation in the direction of the condensate
$\langle\vec\pi\rangle$, while the charged pions are the Nambu--Goldstone bosons corresponding to
the broken generators.}

We may now consider the combined effect of the lattice and QED, which is relevant in the case
that $c_1\sim c_2\sim g^2c_3$, or $m/\Lambda_{\rm QCD}\sim a^2\Lambda_{\rm QCD}^2\sim
g^2\sim e^2$.  If we gauge the complete isospin group, the effective potential is
\begin{equation}
\label{Veffcomb}
V_{\rm eff}+\Delta V_{\rm eff}=-c_1\sigma+(c_2-g^2c_3)\sigma^2\ .
\end{equation}
The only change in comparison with Eq.~(\ref{VeffWilson}) is that the low-energy constant
$c_2$ gets replaced with $c_2-g^2c_3$.   It follows that
for $c_2<0$ the situation is as before, and there is a first-order transition at $c_1=0$.   The
lower bound on the pion mass gets larger by a factor $(|c_2|+g^2c_3)/|c_2|$.
If
$c_2>0$, $c_2-g^2c_3$ flips sign when $a\to 0$, and the Aoki phase thus gets pushed away
from the continuum limit; closer to the continuum limit, when $c_2<g^2c_3$, the first-order scenario applies.

If we only couple electromagnetism, the effective potential is
\begin{equation}
\label{VeffcombEM}
V_{\rm eff}+\Delta V_{\rm eff}=-c_1\sigma+c_2\sigma^2-\frac{1}{2}\,e^2c_3(\sigma^2+\pi_3^2)\ .
\end{equation}
This potential forces $\langle\sigma\rangle^2+\langle\pi_3\rangle^2=1$, and thus any Aoki condensate is now forced into the
third direction.   Isospin is explicitly broken, but parity remains spontaneously broken when
$\langle\pi_3\rangle\ne 0$.   There are no exact Nambu--Goldstone bosons, and inside the Aoki phase
\begin{equation}
\label{masses}
m^2_{\pi^\pm}=e^2c_3/f^2\ ,\qquad m^2_{\pi^0}=2c_2\left(1-\frac{c_1^2}{4c_2^2}\right)/f^2\ .
\end{equation}
Whether the charged or neutral pion mass is larger now depends on the detailed values of
the various couplings.

\section{Staggered fermions}
Next, we consider QCD with two staggered fermions $\omega_i$, $i=1,2$.   In the continuum
limit, this theory has eight flavors, because of the four-fold doubling for each staggered
fermion.\footnote{Often the ``extra'' flavors due to this doubling are referred to as ``taste''
instead of flavor, but here we will refer to all fermions as flavors in the continuum limit.}  We project these staggered fermions onto even and odd sites:
\begin{eqnarray}
\label{projection}
\chi_i(x)&=&\frac{1}{2}(1+\epsilon(x))\omega_i(x)\ ,\qquad\chibar_i(x)=\omegabar_i(x)\frac{1}{2}(1-\epsilon(x))\ ,\\
\lambda_i(x)&=&\frac{1}{2}(1-\epsilon(x))\omega_i(x)\ ,\qquad\lambdabar_i(x)=\omegabar_i(x)\frac{1}{2}(1+\epsilon(x))\ ,\nonumber
\end{eqnarray}
where
\begin{equation}
\label{eps}
\epsilon(x)=(-1)^{x_1+x_2+x_3+x_4}\ .
\end{equation}
The massless theory with this fermion content has an exact $SU(2)_\chi\times SU(2)_\lambda$
flavor symmetry.   In the continuum limit, the lattice fields $\chi_i$ lead to four continuum
Dirac fields $\psi_{1,2,3,4}$, and likewise, the fields $\lambda_i$ yield four continuum Dirac fields
$\psi_{5,6,7,8}$ \cite{DS1983,GS1984}.  It is important to observe that whether any of the exact
lattice symmetries are to be considered as axial symmetries depends on what type of mass
term is added to the theory.   Here we will keep the theory massless, but instead consider the
effect of weakly gauging (some of) the exact flavor symmetries.

On the lattice, dynamical symmetry breaking is expected to take place, and various condensates
may form as a consequence.   One possible condensate is $\sum_{k=1}^8\psibar_k\psi_k$, which is the
continuum limit of
\begin{equation}
\label{onelink}
\sum_{i,\mu}\left(\chibar_i(x)U_\mu(x)\chi_i(x+\mu)+\lambdabar_i(x)U_\mu(x)\lambda_i(x+\mu)
+\mbox{h.c.}\right)\ .
\end{equation}
This operator is invariant under $SU(2)_\chi\times SU(2)_\lambda$; note that an operator such as
$\chibar_i(x)\chi_i(x)$ cannot be constructed because of the projections in Eq.~(\ref{projection}).
The minimum number of links in any lattice
operator with continuum limit $\sum_{k=1}^8\psibar_k\psi_k$ is one.   A different condensate, corresponding to a single-site operator
\begin{equation}
\label{site}
\sum_i\left(\chibar_i(x)\lambda_i(x)+\lambdabar_i(x)\chi_i(x)\right)
\end{equation}
also exists;
in the continuum limit it takes the form
\begin{equation}
\label{sscond}
\psibar_5\psi_1+\psibar_6\psi_2+\psibar_7\psi_3+\psibar_8\psi_4+\mbox{h.c.}\ .
\end{equation}
In contrast to the one-link operator (\ref{onelink}),
this latter condensate breaks $SU(2)_\chi\times SU(2)_\lambda$ down to $SU(2)_{\rm diag}$.   Clearly, these two condensates are not
equivalent on the lattice, even if they are in the continuum.   In the continuum, the flavor group
enlarges to $SU(8)\times SU(8)$, both condensates break this to $SU(8)_{\rm diag}$, and they can be
rotated into each other by the continuum flavor group.   All we need to support this observation is to
assume that the continuum limit of the lattice theory is universal.

The two condensates are no longer equivalent when we gauge some subgroup of the lattice
symmetry group, even in the continuum limit, as we will show through two examples.   First, let us gauge just the $U(1)$ group
 of transformations generated by $T_3^\epsilon$, where
\begin{equation}
\label{T3eps}
T_3^\epsilon= T_3^\chi-T_3^\lambda\ .
\end{equation}
Here $T_a^\chi$ and $T_a^\lambda$ are the generators of $SU(2)_\chi$
and $SU(2)_\lambda$, respectively.  To leading order in the weak gauge coupling $e$,
the effective potential generated by the weak gauge fields is \cite{MP1980}
\begin{equation}
\label{VeffU1eps}
V_{\rm eff}=-e^2C\,\tr(\Sigma Q_R\Sigma^\dagger Q_L)\ ,
\end{equation}
with $\Sigma\in SU(8)$ the non-linear Nambu--Goldstone field,
$C>0$ \cite{EW1983}, and in which $Q_R$ and $Q_L$ are spurion fields.   In order to
represent the coupling to the $U(1)$ gauge field they need to be fixed; to what values depends on the
basis, as we will see below.   Once we choose $Q_R$ and $Q_L$, this potential breaks $SU(8)\times SU(8)$, and the one-link and
single-site condensates are no longer degenerate.
Let us work out in more detail what this implies for the condensate $\langle\Sigma\rangle$.

On the one-link basis, {\it i.e.}, the basis on which the one-link condensate is diagonal in the
continuum limit, the corresponding lattice one-link mass term (\ref{onelink}) does not break the $U(1)$ symmetry.   On this basis, the generator for $U(1)$ transformations takes the form \cite{Golterman:2014lea}\footnote{$I_n$ is the $n\times n$ identity matrix.}
\begin{equation}
\label{T31link}
T_3^\epsilon=\tau_3\times\tau_3\times I_2\ .
\end{equation}
This indeed looks vector-like, and we thus have that
\begin{equation}
\label{VeffU1eps1link}
Q_R=Q_L=T_3^\epsilon\quad\rightarrow\quad V_{\rm eff}=-e^2C\,\tr(\Sigma T_3^\epsilon\Sigma^\dagger T_3^\epsilon)\ ,
\end{equation}
with values $V_{\rm eff}(\Sigma_{\rm 1-link})=-24e^2C$ and $V_{\rm eff}(\Sigma_{\rm site})=+24e^2C$.
Here $\Sigma_{\rm 1-link}=I_8$ corresponds to the continuum condensate
$\sum_{k=1}^8\psibar_k\psi_k$, with lattice version (\ref{onelink}), and $\Sigma_{\rm site}=\tau_1\times I_4$
corresponds to the continuum condensate (\ref{sscond}), with lattice version (\ref{site}).

It is instructive to check this result on a different basis, the single-site basis on which the
single-site condensate is, by definition, diagonal in the continuum limit.\footnote{For the explicit construction of this
basis, see Ref.~\cite{Golterman:2014lea}.}   The basis transformation taking the one-link basis into the single-site basis takes $T_3^\epsilon$ into
\cite{Golterman:2014lea}
\begin{equation}
\label{T3site}
\tilde T_3^\epsilon=-\gamma_5(\tau_2\times I_2\times I_2)\ ,
\end{equation}
where we use tildes to indicate that we are now working on the single-site basis.
This has the appearance of an
axial symmetry; and we have that \begin{equation}
\label{VeffU1eps0link}
Q_R=-Q_L=\tilde T_3^\epsilon\quad\rightarrow\quad V_{\rm eff}=+e^2C\,\tr(\tilde\Sigma\tilde T_3^\epsilon\tilde\Sigma^\dagger\tilde T_3^\epsilon)\ ,
\end{equation}
where in the effective theory the $\gamma_5$ is omitted from $\tilde T_3^\epsilon$.
The minus sign between $Q_R$ and $Q_L$ follows from the presense of the $\gamma_5$ in
Eq.~(\ref{T3site}).   On this basis, on which by definition the single-site condensate takes the form
$\tilde\Sigma=I_8$, it
is straightforward to show that the one-link condensate has flavor structure
$\tilde\Sigma=\tau_3\times I_4$, and the effective potential (\ref{VeffU1eps}) takes the
values $V_{\rm eff}(\tilde\Sigma_{\rm 1-link}=\tau_3\times I_4)=-24e^2C$ and $V_{\rm eff}(\tilde\Sigma_{\rm site}=I_8)=+24e^2C$.    Never mind what basis we use to find the pattern
of symmetry breaking, the result is that the condensate minimizing $V_{\rm eff}$ is the one-link
condensate, which leaves the $U(1)$ unbroken.

As in the previous section, we also consider the example in which we introduce weakly coupled
gauge fields for the whole group $SU(2)_\chi\times SU(2)_\lambda$.   An intriguing possibility
would be that the condensate breaks this group down to a smaller group, producing a dynamical
Higgs mechanism that would render some of the weak gauge boson massive.   Now the
effective potential generated by the weak gauge fields is
\begin{equation}
\label{Vefffull}
V_{\rm eff}=-g_\chi^2C\,\sum_a\tr(\Sigma T^\chi_a\Sigma^\dagger T^\chi_a)
-g_\lambda^2C\,\sum_a\tr(\Sigma T^\lambda_a\Sigma^\dagger T^\lambda_a)\ .
\end{equation}
The low-energy constant $C$ is the same as in Eq.~(\ref{VeffU1eps}).   Working on the
one-link basis, we find that $V_{\rm eff}(\Sigma=I_8)=-12(g_\chi^2+g_\lambda^2)C$,
while $V_{\rm eff}(\Sigma=\tau_1\times I_4)=0$, and again the one-link condensate
$\Sigma=I_8$ is the absolute minimum.   Again, the vacuum alignment mechanism causes
the flavor group $SU(2)_\chi\times SU(2)_\lambda$ to remain unbroken, and no dynamical
Higgs mechanism takes place.   As before, this exercise can be repeated on the single-site
basis, with, of course, the same conclusion.

\section{Conclusion}
We have shown, through examples, how lattice artifacts, quark-mass induced contributions, and
weak interactions can all compete in determining the pattern of symmetry breaking in a strongly
coupled gauge theory.    Of course, in the continuum limit, quark-mass and weak-coupling effects
dominate, but these examples demonstrate that lattice artifacts can mask the correct phase
diagram of the theory.   We find that the mechanism of vacuum alignment prevents
a condensate that would imply a dynamical Higgs mechanism in the weak sector from developing.
This disproves claims to the contrary in the literature \cite{CV2013}.   The only assumption
underlying our conclusions is that of universality.   In the staggered case, universality implies
the equivalency of the single-site and one-link condensates in the continuum limit, in the
absence of the coupling to weak gauge fields.   Close enough to the continuum limit, only
the weak interactions break the global flavor symmetry, and thus only weak interactions
determine the phase diagram.

This type of analysis can be extended to composite Higgs models of interest to BSM physics.
In Ref.~\cite{Golterman:2014yha} we considered for instance the $SU(5)/SO(5)$ coset model
of relevance for the ``Littlest Higgs'' model \cite{LSTH}.   In Ref.~\cite{Golterman:2014lea} we
also consider a staggered example with only six flavors in the continuum limit, by making use
of one normal and one reduced staggered fermion \cite{STW1981,DS1983,GS1984}.
As a corrolary of our discussion, we
observe that in order to determine the low-energy constants $c_3$ in the Wilson case, and
$C$ in the staggered case, weak gauge fields only need to be coupled to flavor symmetries
that are exact on the lattice.   This observation generalizes to other cases of interest (such
as the $SU(5)/SO(5)$ coset model) as well.

\section*{Acknowledgements}
We acknowledge discussions with Simon Catterall.
MG thanks the School of Physics and Astronomy of Tel Aviv University
and YS thanks the Department of Physics and Astronomy of San Francisco
State University for hospitality.
MG is supported in part by the US Department of Energy, and
YS is supported by the Israel Science Foundation under grants no.~423/09 and~449/13.

\end{document}